\documentclass[reprint,superscriptaddress,preprintnumbers,nofootinbib,amsmath,amssymb,aps]{revtex4-2}

\usepackage{CJK}
\usepackage{graphicx}
\usepackage{dcolumn}
\usepackage{bm}

\usepackage{multirow}
\usepackage{booktabs}
\usepackage{enumerate}
\usepackage{amssymb}    
\usepackage{color}         
\usepackage{graphicx}     
\usepackage{mathrsfs} 
\usepackage{hyperref} 
\hypersetup{colorlinks=true,linkcolor=blue,citecolor=blue}
\usepackage{ upgreek } 
\usepackage{color,xcolor,fancybox,epsf,rotating,colordvi}

\usepackage{ulem}

\newcommand{\SLASH}[2]{\makebox[#2ex][l]{$#1$}/}
\newcommand{\eslash}{\SLASH{E}{.2}}

\newcommand{\MeV}{~\rm MeV}
\newcommand{\GeV}{~\rm GeV}
\newcommand{\TeV}{~\rm TeV}
\newcommand{\fbm}{{~\rm fb}^{-1}}
\newcommand{\fb}{~\rm fb}

\begin{document}


\title{
95 GeV light Higgs in the top-pair-associated diphoton channel at the LHC \\ in the minimal dilaton model
}

\author{Kun Wang}
\email[]{kwang@usst.edu.cn}
\affiliation{College of Science, University of Shanghai for Science and Technology, Shanghai 200093, China}

\author{Jingya Zhu}
\email[]{zhujy@henu.edu.cn} 
\affiliation{School of Physics and Electronics, Henan University, Kaifeng 475004, China}

\date{\today}

\begin{abstract}
Motivated by experimental hints and theoretical frameworks indicating the existence of an extended Higgs sector, we explore the feasibility of detecting a 95 GeV light Higgs boson decaying into a diphoton within the minimal dilaton model at the 14 TeV LHC. 
Initially, we identify the correlations between the production cross section, decay branching ratios, and model parameters, e.g., the scalar mixing angle $\sin\theta_S$. 
Subsequently, we utilize Monte Carlo simulations to generate the signal of the light Higgs boson via the $pp \to t\bar{t}(s\to \gamma\gamma)$ process, along with the corresponding backgrounds. 
To effectively separate the signal from the dominant backgrounds $tt\gamma\gamma$, we employ a meticulous cut-based selection process. 
Ultimately, we find that with an integrated luminosity of $L = 3000 \fbm$, the regions of $|\sin\theta_S|>0.2$ can be covered over the $3\sigma$ level.
\end{abstract}

\maketitle
\newpage


\section{\label{sec:Intro}Introduction}
Whether there is another Higgs scalar remains a fundamental and unresolved question, especially following the discovery of the 125 GeV Higgs boson \cite{CMS:2012qbp, ATLAS:2012yve}. 
Beyond the Standard Model (SM), numerous new physics theories naturally suggest the existence of additional Higgs scalars, alongside a $125\GeV$ SM-like Higgs boson.  
In recent years, more and more data have demonstrated that the discovered $125\GeV$ Higgs boson closely resembles the SM, which narrows down the scope of new physics theories, although several theoretical models remain viable.
Direct detection of additional scalar particles, particularly in the low-mass region, is crucial for testing these new physics models. 
Experimentalists have conducted extensive direct searches at colliders such as LEP, Tevatron, and LHC, yielding some intriguing hints.

The first indication of a light Higgs boson emerged in 2003 when the LEP experiment reported a 2.3$\sigma$ local excess in the $e^+ e^- \to Z(H\to b \bar{b})$ channel around 98 GeV \cite{LEPWorkingGroupforHiggsbosonsearches:2003ing}. 
Subsequent LHC experiments provided further clues in 2015, with the CMS collaboration observing a 2.0$\sigma$ local significance of a diphoton excess at around 97 GeV with 8 TeV 19.7 $\fbm$ data \cite{CMS:2015ocq}.
This was followed in 2018 by a 2.8$\sigma$ local (1.3$\sigma$ global) excess in the 95.3 GeV diphoton invariant mass, using a combination of 13 TeV 35.9 $\fbm$ data \cite{CMS:2018cyk}.
In 2022, observations by CMS of a 2.6$\sigma$ local (2.3$\sigma$ global) $\tau^+\tau^-$ excess in the 95-100 GeV range with 13 TeV 138 $\fbm$ data further supported these hints \cite{CMS:2022goy}. 
Most recently, in 2023, CMS and ATLAS reported diphoton excesses around 95 GeV with 2.9$\sigma$ local (1.9$\sigma$ global) \cite{CMS:2023yay} and 1.7$\sigma$ local excesses \cite{ATLAS:2023jzc}, respectively, using 13 TeV 132.2 $\fbm$ and 140 $\fbm$ data.

The recurring excesses around 95 GeV suggest the potential existence of an undiscovered light scalar particle at this mass. 
Numerous recent studies have attempted to account for this 95 GeV excess through various new physics models \cite{
Biekotter:2019kde,Heinemeyer:2021msz,Biekotter:2022jyr,Aguilar-Saavedra:2023tql,Choi:2019yrv,Ma:2020mjz,Li:2022etb,Ellwanger:2023zjc,Dev:2023kzu,Bonilla:2023wok,Liu:2024cbr,Li:2023kbf,Abbas:2023bmm,Ahriche:2023hho,Cao:2023gkc,Ellwanger:2023zjc,Maniatis:2023aww,Belyaev:2023xnv,Azevedo:2023zkg,Bonilla:2023wok,Bhatia:2022ugu,Chen:2023bqr,Ahriche:2023wkj,Arcadi:2023smv,Borah:2023hqw,Banik:2023vxa,Dutta:2023cig,Bhattacharya:2023lmu,Ashanujjaman:2023etj,Biekotter:2023oen,Escribano:2023hxj,Biekotter:2023jld,Banik:2023ecr,Coloretti:2023wng,Ahriche:2022aoj,Benbrik:2022azi,Biekotter:2022abc,Biekotter:2021qbc,Abdelalim:2020xfk,Biekotter:2020cjs,Aguilar-Saavedra:2020wrj,Cao:2019ofo,Kundu:2019nqo,Biekotter:2019kde,Cao:2016uwt,Iguro:2022dok,Iguro:2022fel,Biekotter:2021ovi}. 

We focus on the Minimal Dilaton Model (MDM) \cite{Abe:2012eu, Abe:2013rla, Cao:2013cfa,Ahmed:2015uqt,Ahmed:2019csf}, a straightforward extension of the SM.
The MDM introduces a singlet scalar field $S$ and a fermion top partner, 
which is inspired by concepts from strong interaction theory and topcolor theory.
This model aims to clarify the mechanisms underlying electroweak symmetry breaking and the role of scale invariance at high energy levels. The singlet does not directly couple with fermions and gauge bosons in SM, thus maintaining its status as a pseudo Nambu-Goldstone particle of broken scale invariance.
In scenarios where the top partner has considerable mass, the model essentially converges to the SM plus one scalar. However, a distinct characteristic of the MDM emerges in such contexts: even when the mixing angle between the singlet and the SM-like Higgs boson is zero, the light scalar remains capable of being produced through gluon-gluon fusion processes and subsequently decaying into diphoton channels via loops involving the top partner.

Our previous analyses indicate that the MDM is capable of producing a 95 GeV scalar particle \cite{Liu:2018ryo}. 
In this work, we explore the detectability of this scalar at the LHC. 
To minimize the QCD background, we focus on the production of the $95\GeV$ scalar associated with a top pair, decaying to diphoton, with the top quark decaying into $W^+b$, and the $W$ boson cascade decaying into leptons accompanied by missing transverse energy $\eslash_T$.

The remainder of the paper is structured as follows: 
In Sec.~\ref{sec2}, we introduce the Minimal Dilaton Model and delineate the parameter space under theoretical and experimental constraints. 
In Sec.~\ref{sec3}, we use Madgraph for detailed collider simulations and then discuss the results. 
Finally, in Sec.~\ref{sec4}, we present our conclusions.

\section{The Minimal Dilaton Model}
\label{sec2}
The Minimal Dilaton Model introduces a dilaton-like singlet field, denoted as $S$, and a vector-like fermion field, represented by $T$. The Lagrangian of the model is defined as follows \cite{Abe:2012eu,Abe:2013rla}:
\begin{align}
\mathcal{L} & = \mathcal{L}_{\mathrm{SM}}^{\mathrm{without~}V(H)} \nonumber \\
& -\frac{1}{2} \partial_{\mu} S \partial^{\mu} S   -\bar{T}\left(\SLASH{D}{.5}+\frac{M}{f} S\right) T   \nonumber \\ 
& -\left[y^{\prime} \bar{T}_{R}\left(q_{3 L} \cdot H\right)+\text { h.c. }\right] -\tilde{V}(S, H)  ~,
\end{align}
where $\mathcal{L}_{\mathrm{SM}}^{\mathrm{without~}V(H)}$ represents the SM Lagrangian excluding the scalar potential. 
$M$ is the scale of strong dynamics, and $f$ is the vacuum expectation value (VEV) of the singlet field $S$. 
The term $q_{3 L}$ refers to the quark $SU(2)_L$ doublet of the third generation. 
The modified scalar potential incorporating the scalar $S$, given by $\tilde{V}(S, H)$, is formulated as \cite{Abe:2012eu,Abe:2013rla}:
\begin{align}
\tilde{V}(S, H) & =  m_{H}^{2}|H|^{2}+\frac{\lambda_{H}}{4}|H|^{4}  \nonumber \\
&+ \frac{m_{S}^{2}}{2} S^{2}+\frac{\lambda_{S}}{4 !} S^{4}+\frac{\kappa}{2} S^{2}|H|^{2}   ~,
\end{align}
where $m_{H}$, $\lambda_{H}$, $m_{S}$, $\lambda_{S}$ and $\kappa$ are free real parameters.
For convenience, we introduce a dimensionless parameter \cite{Abe:2012eu,Abe:2013rla}:
\begin{eqnarray}
\eta & = & \frac{v}{f} N_{T} ~.
\end{eqnarray}
Here, $v$ and $f$ represent the VEVs of the Higgs field $H$ and the singlet field $S$, respectively, with $v=246 \GeV$. The variable $N_{T}$ denotes the number of the field $T$, where $N_{T}=1$ in the minimal model. To achieve a larger value of $\eta$, one may increase $N_{T}$, thereby avoiding the necessity for an excessively small $f$.

In the Minimal Dilaton Model, the fields $S$ and $H$ mix, resulting in the formation of two CP-even mass eigenstates: the SM-liked Higgs boson $h$ and the light Higgs $s$ (also called dilaton). The mixing angle, denoted as $\theta_{S}$, is defined by the following equations:
\begin{eqnarray}
H^{0}  & = & \frac{1}{\sqrt{2}}\left(v+h \cos \theta_{S}-s \sin \theta_{S}\right) \, , \\
S  & = & f+h \sin \theta_{S}+s \cos \theta_{S} \, ,
\end{eqnarray}
where $H^{0}$ represents the neutral component of the Higgs field, and $v$ and $f$ are the VEVs of the Higgs and the singlet field, respectively.


Additionally, the mixing angle $\theta_{L}$ for the top quark involving the mass eigenstates $t$ and $t'$ can be defined as follows:
\begin{eqnarray}
q_{3 L}^{u}  & = & \cos \theta_{L} t_{L}+\sin \theta_{L} t_{L}^{\prime}, \\
T_{L}  & = & -\sin \theta_{L} t_{L}+\cos \theta_{L} t_{L}^{\prime} \,.
\end{eqnarray}



In the conditions where $m_{t'} \gg m_{t}$ and $\tan \theta_{L} \ll m_{t'}/m_{t}$, the normalized couplings of the Higgs boson $h$ and the light higgs $s$ can be derived as \cite{Abe:2012eu,Abe:2013rla}:
\begin{eqnarray}
C_{hVV}/SM & = & C_{hff}/SM  =  \cos\theta_S  \, ,\\
C_{sVV}/SM & = & C_{sff}/SM  =  -\sin\theta_S  \, ,
\end{eqnarray}
where $V$ represents either the $W^{\pm}$ or $Z$ boson, and $f$ signifies the fermions, excluding the top quark. 
Here, $C_{hVV}/SM$ and $C_{hff}/{SM}$ represent the normalized couplings of the Higgs boson $h$ to vector bosons and fermions, respectively, relative to their SM predictions. Similarly, $C_{sVV}/{SM}$ and $C_{sff}/{SM}$ denote the normalized couplings of the light Higgs $s$ to vector bosons and fermions. 
The normalized couplings between the light Higgs $s$ and the top quark states $t$ and top partner $t'$ are specified as follows \cite{Abe:2012eu,Abe:2013rla}:
\begin{eqnarray}
\label{eq:cstt}
C_{st\bar{t}}/SM & = &-\cos^2\theta_L\sin\theta_S+\eta\sin^2\theta_L\cos\theta_S \, ,\\
C_{st'\bar{t'}}/SM & = &-\sin^2\theta_L\sin\theta_S+\eta\cos^2\theta_L\cos\theta_S\, .
\end{eqnarray}
Here, the terms $C_{st\bar{t}}/SM$ and $C_{st'\bar{t'}}/SM$ represent the normalized couplings of the light Higgs $s$ to the top quark pair and the top partner pair, respectively.
When the mass of the light Higgs, $m_s$, is set to 95 GeV, the normalized loop-induced couplings with gluons and photons can be approximated as follows \cite{Liu:2018ryo}:
\begin{eqnarray}
C_{sgg}/SM &\approx &-\sin\theta_S+\eta\cos\theta_S \, ,\\
C_{s\gamma\gamma}/SM &\approx& -\sin\theta_S-0.31\eta\cos\theta_S \, .
\end{eqnarray}
Here, $C_{sgg}/SM$ and $C_{s\gamma\gamma}/SM$ denote the normalized couplings of the light Higgs $s$ to gluons and photons, respectively.

The free parameters in this model are $\eta$, $m_s$, $m_{t'}$, $\tan\theta_S$, and $\sin\theta_L$.

\section{Results and Discussions}
\label{sec3}
In this work, we will investigate the possibility of searching for the light Higgs in $pp \to t \bar{t}s$ process at the 14 TeV HL-LHC.
In this model, the mass of the top partner $m_{t'}$ is assumed to be very large to satisfy constraints derived from LHC observations, 
and the mass of the light Higgs $m_{s}$ is fixed at 95 GeV to be confronted with the CMS result.

From our previous studies \cite{Liu:2018ryo}, which aimed to meet experimental and theoretical constraints such as vacuum stability, avoidance of Landau poles, and results from low-mass Higgs or resonance searches at LEP, Tevatron, and LHC, we have carefully examined the parameter space.
Accordingly, we will consider such a parameter space for all constraints and the existence of a 95 GeV light Higgs:
\begin{eqnarray}
0.01<\eta &<&1.2, \quad\left|\tan \theta_{S}\right|<0.75, \quad  \left|\sin \theta_{L}\right|<0.2,  \nonumber \\
m_{s} &=& 95 \mathrm{GeV}, \qquad \qquad m_{t^{\prime}} = 1.67 \mathrm{TeV} . \nonumber
\end{eqnarray}

\begin{figure*}[!htbp]
\includegraphics[width=0.9\textwidth]{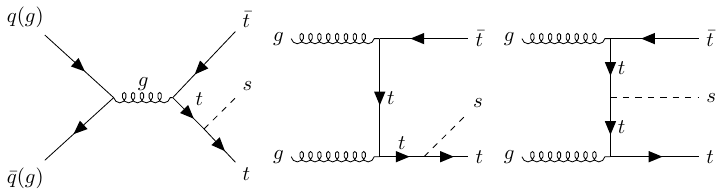}
\caption{\label{fig:feynman}Feynman diagrams for $pp \to t \bar{t} s $ in the Minimal Dilaton Model (MDM) at the tree level.}
\end{figure*}

\begin{figure*}[!htbp]
\includegraphics[width=1\textwidth]{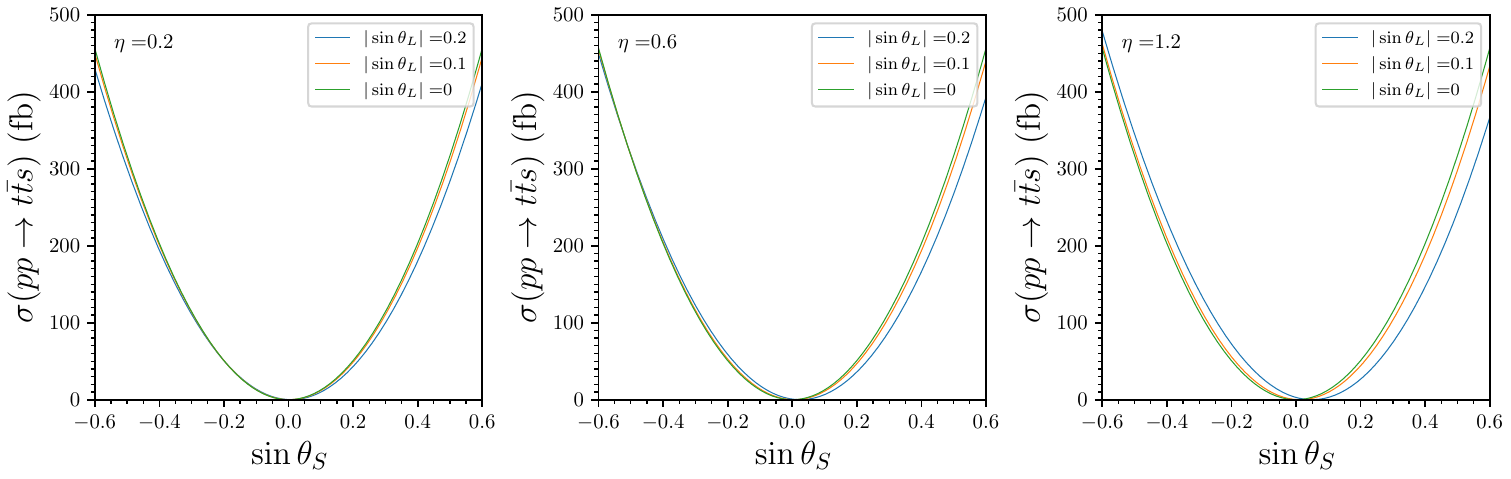}
\caption{\label{fig:1}The cross section of the $p p \to t \bar{t} s$ process versus $\sin\theta_S$, with $\eta = 0.2~(\rm left),~0.6~(middle), ~1.2~(right)$, respectively. The blue, orange, and green lines represent $|\sin\theta_L| = 0.2, 0.1, 0$, respectively. }
\end{figure*}

In the MDM, the tree-level Feynman diagrams for the process $pp \to t\bar{t}s$ at the LHC is illustrated in Fig.\ref{fig:feynman}.
This diagram captures the mechanism of light Higgs $s$ radiation off top quarks, a pivotal process for probing the Higgs sector postulated by the MDM.
The depicted process bears a close resemblance to the well-known $t\bar{t}H$ process, with the primary distinction being the substitution of the Higgs-top coupling $C_{htt}$ with the light Higgs-top coupling $C_{stt}$.
Given this similarity, the cross-section for the $pp \to t\bar{t}s$ process can be derived analogously to that of the $t\bar{t}H$ process. Specifically, the cross-section is calculated by modifying the interaction vertex to incorporate the $C_{stt}$ coupling.
So the cross section can be approximately expressed as follows:
\begin{equation}
\sigma(p p \to t \bar{t} s) = \sigma(p p \to t \bar{t} H)|^{SM}_{m_H=95\GeV} \times (C_{st\bar{t}}/SM)^2  \,.   
\end{equation}
The Next-to-Leading Order (NLO) data for $\sigma(p p \to t \bar{t} H)|^{SM}_{m_H=95\GeV}$ is found to be $1268\fb$ 
 \cite{LHCHiggsCrossSectionWorkingGroup:2011wcg}. 
By applying Eq.(\ref{eq:cstt}), the resulting cross section as a function of $\sin\theta_S$ is illustrated in Fig.\ref{fig:1}. 
It is evident that the cross section $\sigma(p p \to t \bar{t} s)$ primarily depends on $\sin\theta_S$, and the sign of $\sin\theta_L$ does not in fact affect the result. 
This dependency arises because $|\sin\theta_L| \ll 1$; in Eq.(\ref{eq:cstt}), the term $\sin^2\theta_L$ can be neglected. 
Consequently, the coupling ratio $(C_{st\bar{t}}/SM)^2$ approximates to $\cos^4\theta_L\sin^2\theta_S$, showcasing a quadratic relationship with $\sin\theta_S$.

The light Higgs is mainly decay into $b\bar{b}$, $c\bar{c}$, $\tau^+\tau^-$, $W W^*$, $ZZ^*$, $gg$, $\gamma\gamma$.
The light Higgs decay branching ratio can be calculated from the SM Higgs decay information:
\begin{eqnarray}
    Br_{s\rightarrow xx}=Br_{s\rightarrow xx}^{SM}\times|C_{s xx}/SM|^{2}\times\frac{\Gamma_{tot}^{SM}}{\Gamma_{tot}^{\phi}},
\end{eqnarray}
where the $xx$ represent the $b\bar{b}$, $c\bar{c}$, $\tau^+\tau^-$, $W W^*$, $ZZ^*$, $gg$ and  $\gamma\gamma$, and the $\phi = h, s $ in MDM.
$\Gamma_{tot}^{SM}$ and $\Gamma_{tot}^{\phi}$ are the total decay widths of the 95 GeV SM Higgs $H$ and the light Higgs $s$, respectively.
And the $\Gamma_{tot}^{\phi}$ can be writen as:
\begin{eqnarray}
\Gamma_{tot}^{\phi}=\Gamma_{tot}^{SM}\times\sum_{xx} \left[Br_{s\rightarrow xx}^{SM}\times|C_{s xx}/SM|^{2} \right]
\end{eqnarray}
where the 95 GeV SM Higgs decay information is given as \cite{LHCHiggsCrossSectionWorkingGroup:2011wcg}:
\begin{align*}
Br_{s\rightarrow b\bar{b}}^{SM} &= 0.804,& 
Br_{s\rightarrow c\bar{c}}^{SM} &= 0.0373,& \\
Br_{s\rightarrow \tau^+\tau^-}^{SM} &= 0.0841,&
Br_{s\rightarrow W W^*}^{SM} &= 0.00472,& \\
Br_{s\rightarrow ZZ^*}^{SM} &= 0.000672,& 
Br_{s\rightarrow gg}^{SM} &= 0.0674,& \\
Br_{s\rightarrow \gamma\gamma}^{SM} &= 0.00140,& 
\Gamma_{tot}^{SM}&= 2.32\MeV&
\end{align*}

\begin{figure*}[!htbp]
\includegraphics[width=1\textwidth]{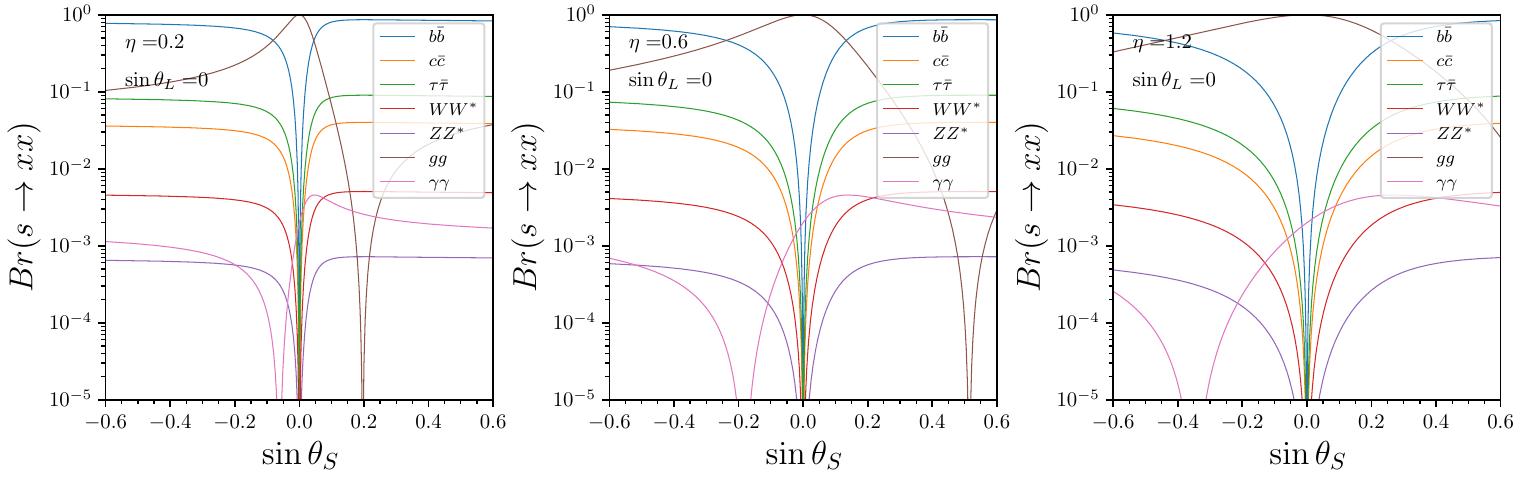} 
\caption{\label{fig:2}Decay branching ratios of the light Higgs boson $s$ as a function of $\sin\theta_S$ in the MDM, with $\sin\theta_L = 0$ and $\eta = 0.2~(\rm left),~0.6~(middle), ~1.2~(right)$, respectively.
The decay channels $xx$ include $b\bar{b}$, $c\bar{c}$, $\tau^+\tau^-$, $W W^*$, $ZZ^*$, $gg$, and $\gamma\gamma$.}
\end{figure*}

\begin{figure*}[!htbp]
\includegraphics[width=1\textwidth]{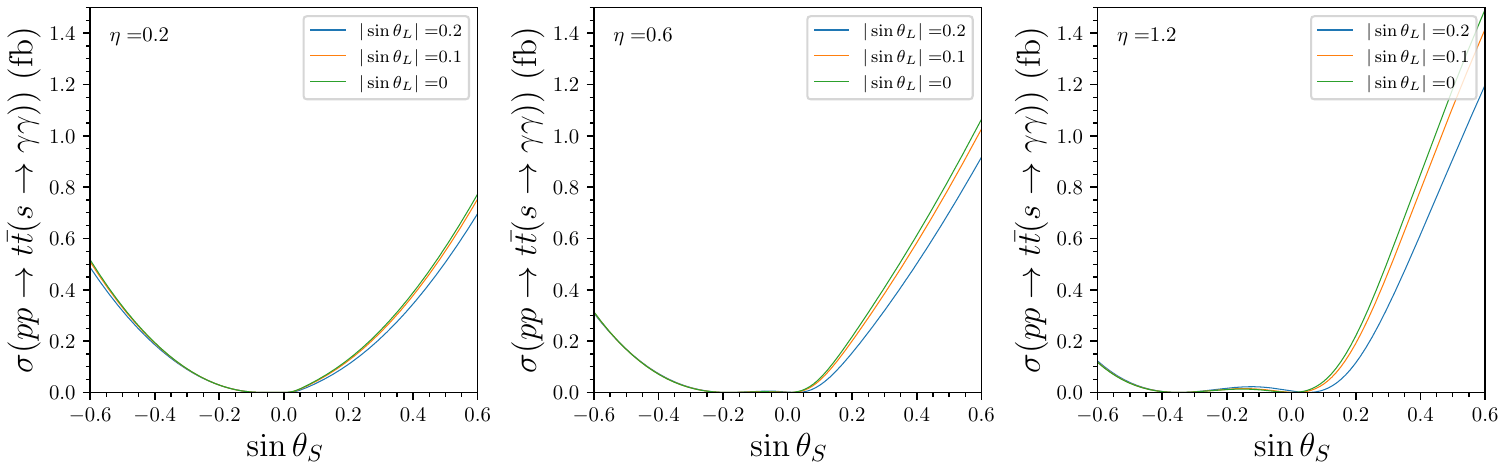} 
\caption{\label{fig:3}The cross section of the $p p \to t \bar{t} (s \to \gamma \gamma)$ process versus $\sin\theta_S$, with $\eta = 0.2~(\rm left),~0.6~(middle), ~1.2~(right)$, respectively
The blue, orange, and green curves represent $|\sin\theta_L| = 0.2, 0.1, 0$, respectively. }
\end{figure*}

In Fig.~\ref{fig:2}, we plot the decay branching ratios of the light Higgs boson $s$ as a function of $\sin\theta_S$. 
Our analysis confirms that the decay branching ratios are not significantly influenced by the $\sin\theta_L$ parameter; therefore, we set $\sin\theta_L=0$.
It is observed that for samples where $\sin\theta_S>0$, the branching ratio $Br(s\to\gamma\gamma)$ is greater than for those with $\sin\theta_S<0$.
Additionally, the minimum value of $Br(s\to\gamma\gamma)$ is influenced by the parameter $\eta$.
Specifically, when $\eta$ is set to 0.2, 0.6, and 1.2, the minimum values of $Br(s\to\gamma\gamma)$ correspond to $\sin\theta_S$ values of 0.1, 0.2, and 0.4, respectively.

In Fig.~\ref{fig:3}, we present the cross section of the $pp \to t\bar{t}(s \to \gamma\gamma)$ process as a function of $\sin\theta_S$.
The cross section $\sigma(pp \to t\bar{t}(s \to \gamma\gamma))$ is calculated as the product of $\sigma(pp \to t\bar{t}s)$ and $Br(s\to\gamma\gamma)$, given by:
\begin{eqnarray}
\sigma(p p \to t \bar{t} (s \to \gamma \gamma)) &=& |\sigma(p p \to t \bar{t} H)|^{SM}_{m_H=95\GeV} \nonumber \\ 
&\times& (C_{st\bar{t}}/SM)^2 
\times  Br_{s\rightarrow \gamma\gamma} 
\end{eqnarray}
Our analysis indicates that the cross section $\sigma(pp \to t\bar{t}(s \to \gamma\gamma))$ is primarily affected by $\sin\theta_S$ and $\eta$, with the $\sin\theta_L$ parameter exerting minimal influence.
Notably, the cross section becomes significantly small when $\sin\theta_S \approx 0$, attributable to $\sigma(pp \to t\bar{t}s) \approx 0$ under such conditions.
Furthermore, for samples where $\sin\theta_S<0$, the cross section $\sigma(pp \to t\bar{t}(s \to \gamma\gamma))$ is smaller compared to those with $\sin\theta_S>0$, due to the correlation of $\sin\theta_S<0$ with lower $Br(s\to\gamma\gamma)$ values.

To assess observability, we conduct collider Monte Carlo simulations to explore the sensitivity of the $pp \to t\bar{t}s$ process at the 14 TeV LHC, focusing on the decay channel:
\begin{equation}
pp \to t\bar{t}s \to W^+ b W^- \bar{b} \gamma\gamma
\end{equation}
At tree level, the signal produces two top quarks and a light Higgs boson in the final state. The top quarks predominantly decay into $b$ quarks and $W$ bosons. Considering the light Higgs boson's decay into a diphoton, the final state consists of two $b$ quarks, two $W$ bosons, and two photons ($WWbb\gamma\gamma$), manifesting as a narrow diphoton resonance centered at the light Higgs mass of 95 GeV.

Backgrounds can be categorized based on their behavior in diphoton systems as either resonant or non-resonant. 
The resonant backgrounds, including $tth$ and $tjh$, feature a Higgs boson that can decay into a diphoton in the final state, with the top quark decaying into a $b$ quark and a $W$ boson. 
In the case of $tjh$, the light jet can be misidentified as a $b$ jet. 
The non-resonant backgrounds consist of processes such as $bb\gamma\gamma$, $tt\gamma\gamma$, $tj\gamma\gamma$, $tt\gamma$, and $Wjj\gamma\gamma$, with $tj\gamma\gamma$ and $Wjj\gamma\gamma$ involving a light jet being misidentified as a $b$ jet.

Signal and background events at the parton level were generated using \texttt{MadGraph5\_aMC@NLO\_v3.4.1} \cite{Alwall:2014hca}. 
Subsequently, \texttt{PYTHIA\_v8.2} \cite{Sjostrand:2014zea} was employed for particle decay, parton shower, and hadronization processes. 
The simulation of the detector response was conducted with \texttt{Delphes\_v3.4.1} \cite{deFavereau:2013fsa}, adopting the default values for mistagging efficiencies. 
Jet clustering was performed with the \texttt{anti-kT} algorithm \cite{Cacciari:2008gp}, and the \texttt{NNPDF23LO1} \cite{NNPDF:2014otw} parton distribution functions were selected for our simulation.
Event analysis was carried out with \texttt{MadAnalysis5\_v1.9.60} \cite{Conte:2012fm}.

\begin{figure*}[!htbp]
\includegraphics[width=1\textwidth]{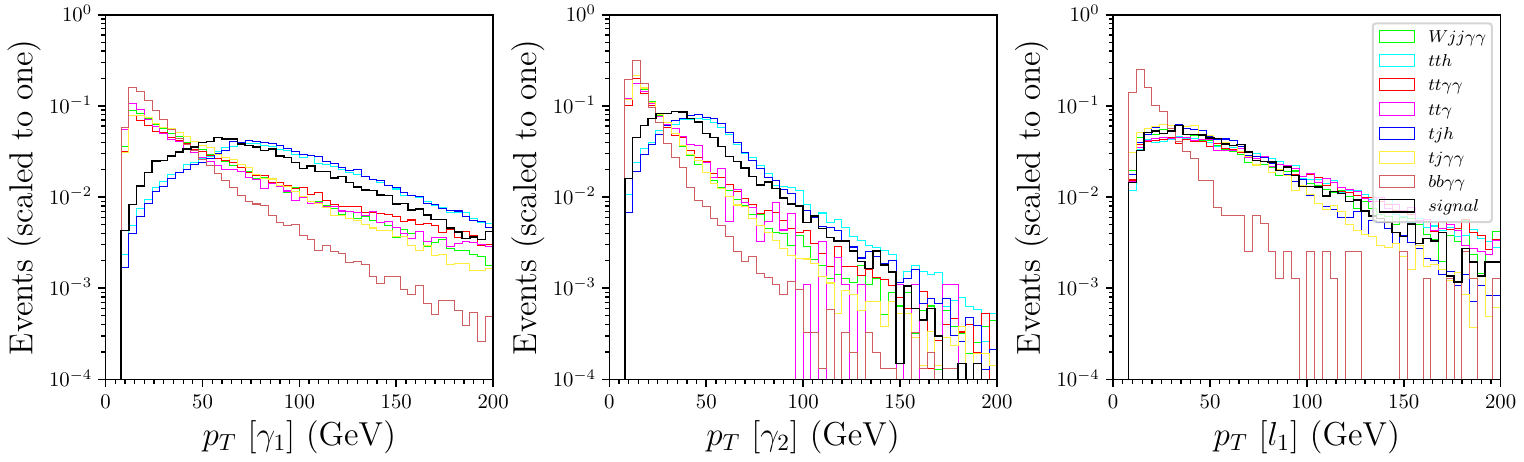} \\
\vspace{10mm}
\includegraphics[width=1\textwidth]{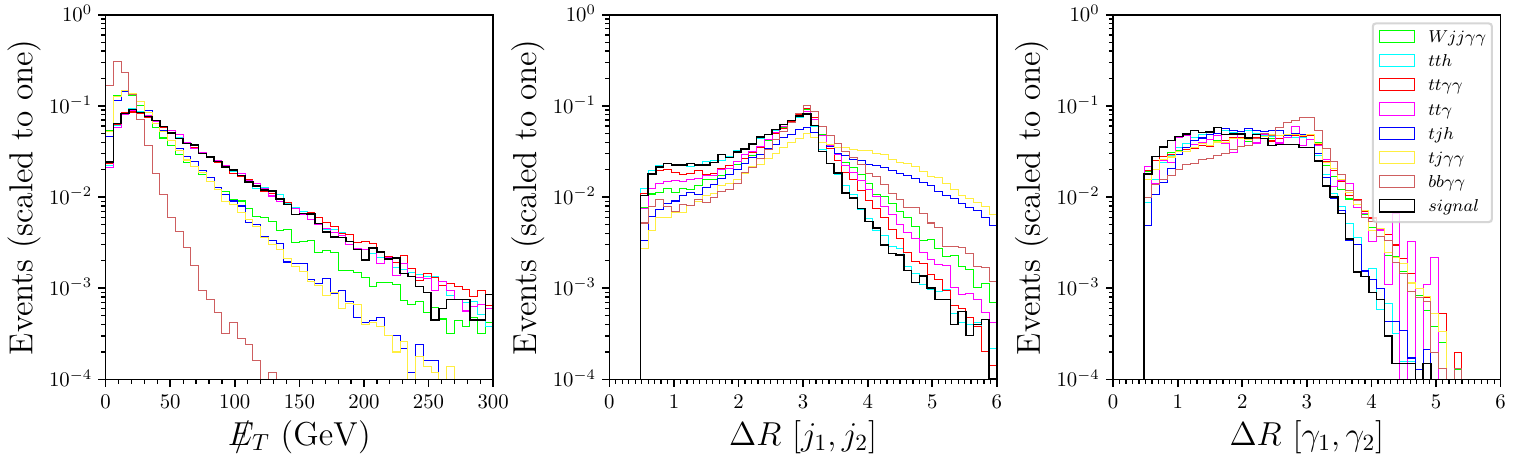}\\
\vspace{10mm}
\includegraphics[width=1\textwidth]{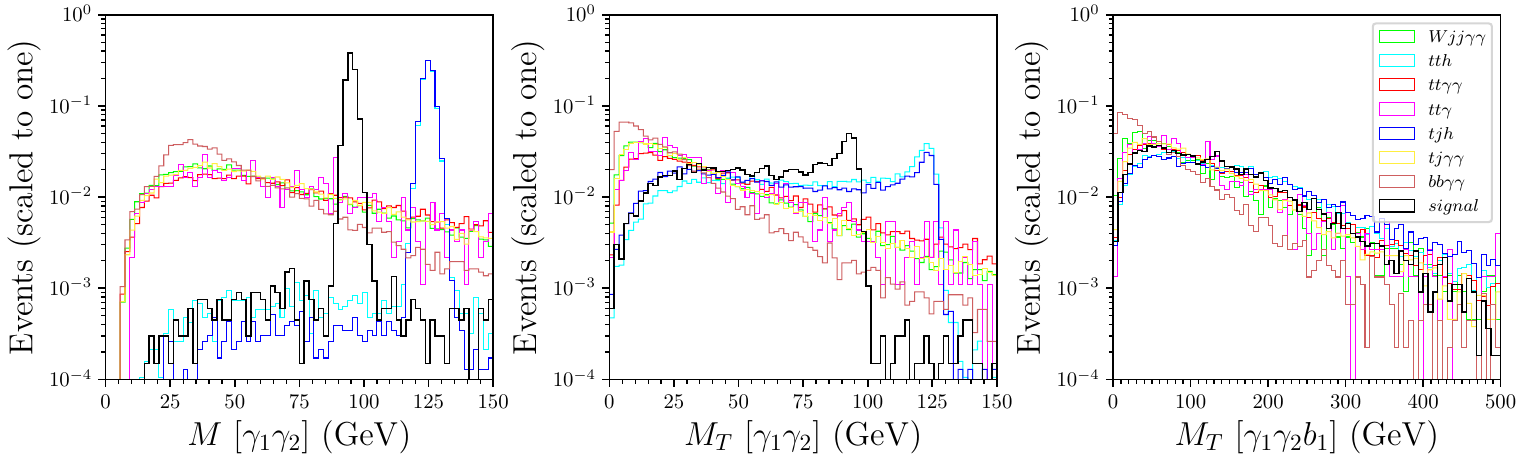} \\
\caption{\label{fig:4}The normalized distributions of nine kinematic variables, including $p_T^{\gamma_1}$ (upper left), $p_T^{\gamma_2}$ (upper middle), $p_T^{\ell_1}$ (upper right), ${\eslash_T}$ (middle left), $\Delta R_{j_1,j_2}$ (middle), $ \Delta R_{\gamma_1, \gamma_2}$ (middle right), 
 $M [\gamma_1\gamma_2]$ (lower left), $M_T[\gamma_1\gamma_2]$ (lower middle), and $M_T[\gamma_1\gamma_2 b_1]$ (lower right), for the signals and SM backgrounds at the 14 TeV LHC.}
\end{figure*}

To determine appropriate kinematic cuts, Fig.~\ref{fig:4} displays normalized distributions for both the signals and SM backgrounds. 
The selection criteria for particles and jets are based on quality metrics for leptons, jets, and photons, specified as follows:
\begin{gather}
 p_{T}^{j/b} > 20 \,  \text{GeV}, \quad p_{T}^{\ell/\gamma} > 10  \, \text{GeV},   \\
|\eta_{i}| < 5,\quad \Delta R_{ij} > 0.4 , (i, j = j, b, \ell, \gamma)   
\end{gather}

To enhance the signal-to-background ratio for the $WWbb\gamma\gamma$ signal amidst the QCD background, it is crucial to include at least one leptonic decay of a $W$ boson into a lepton and a neutrino, manifesting as a lepton with missing transverse energy ($\eslash_T$). To this end, we apply a series of cuts in four stages:
\begin{enumerate}[(1)]
\item \textbf{Basic Cut:} Selection criteria include at least one lepton ($N(\ell) \geq 1$), two photons ($N(\gamma) \geq 2$), and one $b$-jet ($N(b) \geq 1$), alongside thresholds for missing transverse energy and spatial separation:
\begin{equation}
{\eslash}_T > 20 \, \text{GeV}, \quad \Delta R_{j_1,j_2} < 3.1 \, \quad \Delta R_{\gamma_1, \gamma_2} < 3.3
\end{equation}
with specified $p_T$ criteria for photons and leptons:
\begin{equation}
p_{T}^{\gamma_1} > 55 \, \text{GeV}, \quad p_{T}^{\gamma_2} > 25 \, \text{GeV} , \quad 10 < p_{T}^{\ell_1} <100 \, \text{GeV}
\end{equation}

\item \textbf{Diphoton Invariant Mass Cut:} Concentrating on the diphoton invariant mass narrowing down around the light Higgs mass ($95\GeV$) to improve signal clarity:
\begin{equation}
 90 < M [\gamma_1\gamma_2] <100 \, \text{GeV}
\end{equation}

\item \textbf{Transverse Invariant Mass of Diphoton Cut:} Refining the selection further based on the diphoton transverse invariant mass to diminish background:
\begin{equation}
 20 < M_T[\gamma_1\gamma_2] <100 \, \text{GeV}
\end{equation}

\item \textbf{Transverse Invariant Mass of Diphoton with a $b$-Jet Cut:} Narrowing down events based on the combined transverse invariant mass of the diphoton plus a $b$-jet:
\begin{equation}
 40 < M_T[\gamma_1\gamma_2 b] <300 \, \text{GeV}
\end{equation}
\end{enumerate}








\begin{table*}[!htbp]
\centering
\caption{\label{tab:1}The cut flow of the cross sections, for a benchmark point with $\eta=0.2, \sin\theta_S=-0.6, \sin\theta_L = 0.15$.   }
\begin{tabular}{ 
    >{\centering\arraybackslash}p{2.0cm} 
    >{\centering\arraybackslash}p{3.6cm} 
    >{\centering\arraybackslash}p{1.4cm} 
    >{\centering\arraybackslash}p{1.4cm} 
    >{\centering\arraybackslash}p{1.4cm} 
    >{\centering\arraybackslash}p{1.4cm} 
    >{\centering\arraybackslash}p{1.4cm} 
    >{\centering\arraybackslash}p{1.4cm} 
    >{\centering\arraybackslash}p{1.4cm} 
}
\toprule
\multirow{2}{*}{Cuts} &  Signal$(\sigma \times 10^{-3} \fb)$  & \multicolumn{7}{c}{Background$(\sigma\times 10^{-3} \fb )$}                             \\ \cmidrule(r){2-2} 
\cmidrule(l){3-9} 
                      & $tts$    & $bb\gamma\gamma$    & $tj\gamma\gamma$  & $tjh$  & $tt\gamma$     & $tt\gamma\gamma$   & $tth$   & $Wjj\gamma\gamma$   \\ \midrule
Initial             & 500    & 7646000 & 17510 & 78   & 5120000 & 11180  & 563   & 218000 \\
Basic cuts                & 15.65  & 0.00    & 35.02 & 0.97 & 1751.04 & 116.94 & 22.70 & 152.60 \\
$M[\gamma_1 \gamma_2]$                   & 15.00  & 0.00    & 1.40  & 0.00 & 102.91  & 6.48   & 0.02  & 8.72   \\
$M_T[\gamma_1 \gamma_2]$                  & 14.40  & 0.00    & 1.40  & 0.00 & 0.00    & 5.14   & 0.02  & 4.36   \\
$M_T[\gamma_1 \gamma_2 b]$                 & 12.25  & 0.00    & 0.70  & 0.00 & 0.00    & 4.25   & 0.02  & 0.00   \\ \bottomrule
\end{tabular}
\end{table*}

We apply these cuts to our benchmark points and backgrounds to assess the signal efficiency. 
The outcomes of these sequential cuts are presented in Table~\ref{tab:1}. 
Notably, the predominant background originates from the $tt\gamma\gamma$ process. 
After implementing the basic cut, the $bb\gamma\gamma$ background becomes negligible, attributed to the inclusion of a lepton and missing transverse energy ($\eslash_T$) criteria.
Upon applying the cut on the invariant mass around 95 GeV, backgrounds from SM Higgs processes, specifically $tjh$ and $tth$, are effectively eliminated. 
The subsequent cut on the transverse invariant mass of the diphoton system further excludes the background involving missed tagged photons, namely $tt\gamma$. 
Finally, the cut on the transverse invariant mass of the diphoton system with a $b$-jet successfully removes the incorrectly tagged $b$-jet background from the $Wjj\gamma\gamma$ process.

\begin{figure*}[!htbp]
\includegraphics[width=1\textwidth]{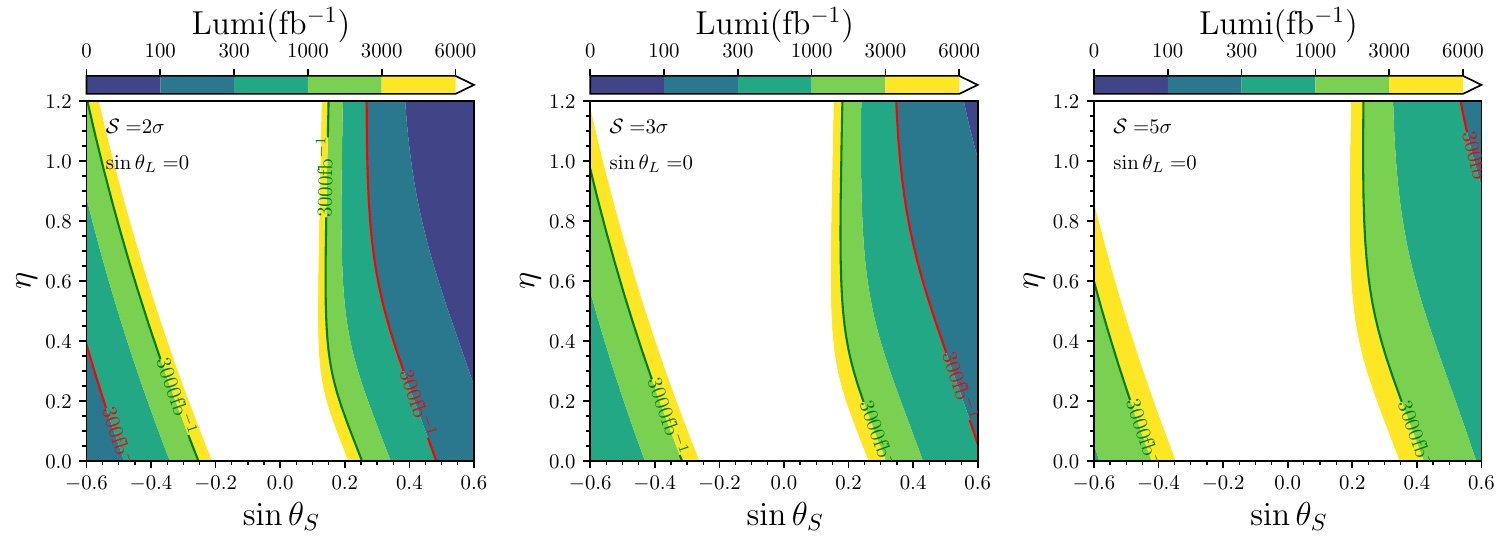} 
\caption{\label{fig:5} Required luminosity to achieve specified statistical significance of $2\sigma$ (left), $3\sigma$ (middle), and $5\sigma$ (right), respectively, on the $\eta$ versus $\sin\theta_S$ plane, with $\sin\theta_L = 0$. The red curve represents the luminosity threshold of $L = 300 \, \text{fb}^{-1}$, and the green curve denotes $L = 3000 \, \text{fb}^{-1}$. }
\end{figure*}

In Fig.~\ref{fig:5}, we illustrate the required luminosity to achieve specified statistical significance levels on the $\eta$ versus $\sin\theta_S$ plane, assuming $\sin\theta_L = 0$. Our analysis confirms that $\sin\theta_L$ has a minimal impact on the outcomes. To estimate the signal significance, we employ the Poisson formula \cite{Cowan:2010js}:
\begin{equation}
\mathcal{S} = \sqrt{2L\left[(S+B) \ln \left(1+\frac{S}{B}\right)-S\right]} \, ,
\end{equation}
where $S$ and $B$ represent the signal and background cross sections, respectively, and $L$ is the integrated luminosity. The results demonstrate how the statistical significance ($\mathcal{S}$) values vary with integrated luminosity $L$. Statistical significance levels of $2\sigma$, $3\sigma$, and $5\sigma$ are illustrated from left to right, respectively. The red line indicates a luminosity threshold of $L = 300 \, \text{fb}^{-1}$, while the green line marks $L = 3000 \, \text{fb}^{-1}$.
From the left figure, it is evident that regions where $\eta+4\sin\theta_S<-2$ or $\eta+4\sin\theta_S>2$ can be covered at the $2\sigma$ level with an integrated luminosity of $L = 300 \, \text{fb}^{-1}$. 
Similarly, regions where $\eta+4\sin\theta_S<-1.2$ or $\sin\theta_S>0.2$ can be covered at the $2\sigma$ level with $L = 3000 \, \text{fb}^{-1}$.
In the middle plane, the region where $\sin\theta_S>0.4$ can reach a $3\sigma$ significance with $L = 300 \, \text{fb}^{-1}$. 
Additionally, areas defined by $\eta+3.3\sin\theta_S<-1$ or $\sin\theta_S>0.2$ can achieve a $3\sigma$ level with $L = 3000 \, \text{fb}^{-1}$.
The right plane demonstrates that regions satisfying $\eta+3\sin\theta_S<-1.2$ or $\sin\theta_S>0.25$ can achieve a $5\sigma$ level at an integrated luminosity of $L = 3000 \, \text{fb}^{-1}$.

\section{Conclusion}
\label{sec4}

In this study, we explore the detection prospects of a $95 \GeV$ light Higgs boson in the Minimal Dilaton Model through the top-pair-associated diphoton process $pp \rightarrow t\bar{t}s (\rightarrow \gamma\gamma)$ at the $14 \TeV$ LHC. 
Monte Carlo simulations have been conducted to analyze the signal and background. 
The final signal state considered is $WWbb\gamma\gamma$, necessitating at least one leptonic decay of $W$ boson to reduce the QCD background.
The production cross section $\sigma(pp \rightarrow t\bar{t}s)$ demonstrates a quadratic relationship with $\sin\theta_S$. 
It's noteworthy that the light Higgs decay branching ratio $Br(s \rightarrow \gamma\gamma)$ is lower for $\sin\theta_S < 0$ compared with that of $\sin\theta_S > 0$. 
Following a meticulous cut-based selection process, we identified that the predominant background is attributed to the SM $tt\gamma\gamma$ process.

Our results indicate that areas where $\eta+4\sin\theta_S < -2$ or $\eta+4\sin\theta_S > 2$ can be covered at the $2\sigma$ level with an integrated luminosity of $L = 300 \, \text{fb}^{-1}$. Similarly, regions defined by $\eta+4\sin\theta_S < -1.2$ or $\sin\theta_S > 0.2$ can be covered at the $2\sigma$ level with $L = 3000 \, \text{fb}^{-1}$. 
Moreover, regions of $\eta+3\sin\theta_S < -1.2$ or $\sin\theta_S > 0.25$ can reach a significance level of $5\sigma$ with an integrated luminosity of $L = 3000 \, \text{fb}^{-1}$.

Overall, for areas where  $|\sin\theta_S|>0.2$, indicative of significant Higgs-dilaton mixing, exclusion or detection at future colliders is feasible with luminosities of $L = 300 \, \text{fb}^{-1}$ or $L = 3000 \, \text{fb}^{-1}$.
Conversely, the detection of areas where $|\sin\theta_S|<0.2$, indicative of minimal Higgs-dilaton mixing, proves challenging to detect through this process, even at an integrated luminosity of $L = 3000 \, \text{fb}^{-1}$.
We aim to further investigate this parameter space in future research, focusing on strategies to identify such scenarios at upcoming collider experiments.

\begin{acknowledgments}
This work was supported by the National Natural Science Foundation of China (NNSFC) under
grant Nos. 12275066 and 11605123.
\end{acknowledgments}

\appendix




\bibliographystyle{apsrev4-1}
\bibliography{apssamp}

\end{document}